\documentclass[11pt]{report}
\usepackage{graphics}
\usepackage{amsfonts}
\usepackage{amssymb}
\voffset - 1.0in
\title{COMMENT ON ``PHASE TRANSITION-LIKE BEHAVIOR IN A LOW-PASS FILTER''}
\author{JACK L. URETSKY \, \, High Energy Physics Division,
Argonne National Laboratories}
\date{\today}
\begin{document}
\maketitle
\indent Krivine and Lesne use an example taken from The Feynman
Lectures\cite{Feyn} in an attempt to illustrate that ``many interesting physical
properties can however be missed because of the improper use of mathematical
techniques''.\cite{paper}  The supposedly incorrect mathematical procedure has
to do with an ordering of limits.  An infinite series that is convergent in the
presence of a small parameter no longer converges when the parameter is set to
zero before the series is summed.

\indent The authors, correctly in my view, emphasize the physical importance of
distinguishing between infinite systems and large finite systems.  In their
example discontinuities in certain physical quantities only exist
(mathematically) for infinite systems.

\indent  I suggest, however, that the authors have demonstrated a different
mathematical point than the one that they propose: infinite series live a life
of their own and need not be constrained to be the limit of sequences of finite
series.  This point was made long ago by Borel and was probably known to Abel
and Cauchy\cite{Borel}.  I emphasize the point with an example of an infinite
series of resistive elements that sum to a negative resistance.  The infinite
series represents different physics from any of the possible
finite series.

\indent Let $\{R_{i}\}$ be a set of resistors, each having resistance
$R_{i}=p^{i}R \ p>1$ and $R$ an arbitrary resistance value.  Then $Z_{n} =
\sum_{i=0}^{n}R_{i}$ is the resistance of a set of such resistors connected in
series, and the value of $Z_{n}$ grows without bound as $n$ increases.  Clearly,
a quantity $Z$ defined by $Z \equiv \sum_{n=0}^{\infty }R_{n}$ makes no sense as
a limit of a convergent sequence of finite sums.

\indent We may, however, emulate Feynman\cite{Feyn} and define $Z$ from the
recursive relation $Z-R=pZ$ which follows from the definition of $Z$ and the
fact that an infinite series less a finite set of its members is still an
infinite series\footnote{The series, in fact, satisfies Hardy's criteria of
${\mathfrak  X}$-summability\cite{Hardy}}.  Solving the last equation for $Z$
leads to the result that
\begin{equation}
Z= -R/(p-1)
\label{Rneg}
\end{equation}
a negative resistance.

\indent Feynman\cite{Feyn} also shows us how to build such infinite-series
resistors.  One simply terminates a finite-series resistor having resistance
$Z_{n}$ with a negative resistance having resistance $ -p^{n+1}R/(p-1)$.  Each
such resistor will then have negative resistance $Z$.

\indent When the quantity $p$ has values $p<1$, there is no difference between
the limit of a sequence $Z_{n}$ of increasing $n$ and the value $Z$ obtained in
Eq. \ref{Rneg}.  This does not mean that Eq. \ref{Rneg} is wrong, as the authors
of Ref. 2 seem to imply.  It does mean that the infinite sum involved represents
two different physical situations when $p<1$ and when $p>1$, involving,
respectively passive and active circuit elements.

\indent This Comment is intended, however, to emphasize the mathematical fact
that infinite (and infinitesimal) mathematical operations may be justified
independently of arguments involving limits.\cite{unlimits}

\indent  I am indebted to Cosmas Zachos for bringing the Borel reference to my
attention.  This work was supported by the U.S. Department of Energy, Division
of High Energy Physics, Contract W-31-109-ENG-38.

\end{document}